\documentclass[12pt,preprint]{aastex}
\usepackage{graphicx,times}
\usepackage{natbib}

\slugcomment{2014.4 v2}

\shorttitle{Pulsar wind model for the spindown of intermittent pulsars}
\shortauthors{Li, Tong, Yan et al.}

\begin{document}

\title{Pulsar wind model for the spin-down behavior of intermittent pulsars}

\author{L. Li\altaffilmark{1,2,3}, H. Tong\altaffilmark{1}, W. M. Yan\altaffilmark{1}, J. P. Yuan\altaffilmark{1},
R. X. Xu\altaffilmark{4,5}, and N. Wang\altaffilmark{1}}

\altaffiltext{1}{Xinjiang Astronomical Observatory, Chinese Academy of Sciences, Urumqi, Xinjiang 830011, China;\\
\it{H. Tong: tonghao@xao.ac.cn}}
    \altaffiltext{2}{University of Chinese Academy of Sciences, 19A Yuquan Road, Beijing, China}
    \altaffiltext{3}{School of Physics, Xinjiang University, Urumqi, Xinjiang, China}
    \altaffiltext{4}{School of Physics, Peking University, Beijing, China}
    \altaffiltext{5}{Kavli Institute for Astronomy and Astrophysics, Peking University, Beijing, China}

\begin{abstract}
 Intermittent pulsars are part-time radio pulsars. They have higher slow down
 rate in the on state (radio-loud) than in the off state (radio-quiet). This gives the evidence
 that particle wind may play an important role in pulsar spindown. The effect of
 particle acceleration is included in modeling the rotational energy loss rate
 of the neutron star.
 Applying the pulsar wind model to the three intermittent pulsars
 (PSR B1931$+$24, PSR J1841$-$0500, and PSR J1832$+$0029),
 their magnetic field and inclination angle are calculated simultaneously.
 The theoretical braking indices of intermittent pulsars are also given.
  In the pulsar wind model, the density of the particle wind
 can always be the Goldreich-Julian density. This may ensure that
 different on states of intermittent pulsars are stable.
 The duty cycle of particle wind can be determined from timing observations.
 It is consistent with the duty cycle of the on state.
 Inclination angle and braking index observations of intermittent pulsars
 may help to test different models of particle acceleration.
 At present, the inverse Compton scattering induced space charge limited flow
 with field saturation model can be ruled out.
 \end{abstract}


\section{Introduction}
Intermittent pulsars are a special class of part-time radio pulsars (Kramer
et al. 2006; McLaughlin et al. 2006; Camilo et al. 2006; Wang et al. 2007).
They have been observed long time scales nulling.
They emit radiation switching between on state and off state.
PSR B1931$+$24 (the first intermittent pulsar, Kramer et al. 2006) exhibits
the alternation of on state (5$\sim$10 d) and off state
(25$\sim$35 d), which is about 38-day quasi-periodicity (Young et al. 2013).
Thereafter, Camilo et al. (2012) and Lorimer et al. (2012) successively
presented extraordinary long term off state of PSR J1841$-$0500 and PSR J1832$+$0029
between their respective on states. More significantly, the spindown rate
of the intermittent pulsar during
the on state is larger than that during the off state. For PSR B1931$+$24,
it enhances $\sim$50\% during its on state compared to its off state. For
PSR J1841$-$0500 and PSR J1832$+$0029, the enhancement is around 150\% and 77\%,
respectively.

Many scenarios have been proposed for intermittent pulsars (mainly centered on PSR B1931$+$24), e.g.,
pulsar with orbiting objects (Li 2006), reactivated
dead pulsars or nulling pulsars viewed at the opposite direction (Zhang et al. 2007),
non-radial oscillation (Rosen et al. 2011), precession (Jones 2012).
However, it is still a puzzle that what is account for the on/off transition
and why the rotation slows down faster when the pulsar is on than when it is off.

An effective model is that an additional particle flow
slows down the rotation when the pulsar is on (Kramer et al. 2006).
A magnetospheric model is presented by Li et al. (2012b) for the spindown of
intermittent pulsars.
But the particle acceleration is not considered in these models
(the magnetospheric model considered the particle acceleration by introducing
a finite conductivity).
Particle acceleration is crucial for the generation of pulsar radio (and high energy) emission.
The pulsar wind model of Xu \& Qiao (2001) considered
the effect of particle wind plus magnetic dipole braking. Different particle
acceleration models are considered there.
This pulsar wind model is applied to the three intermittent pulsars,
i.e., PSR B1931+24, PSR J1841$-$0500, and PSR J1832+0029.
The spindown ratio between the on and off state depends on the particle acceleration potential
and the magnetic inclination angle.
For each source, their magnetic field and inclination angle in different acceleration potentials
are calculated.
Meanwhile, the theoretical braking indices for them are also shown.

Rotational energy loss rate in the presence of a particle wind is calculated in Section 2.
Applying the pulsar wind model to intermittent pulsars is given in Section 3.
Discussions and conclusions are presented in Section 4.

\section{Rotational energy loss rate in the presence of a particle wind}

Pulsars may be viewed as an oblique rotating dipole. The magnetic moment has both
a parallel component and a perpendicular component relative to the rotation axis.
The rotational energy loss rate due to the perpendicular magnetic moment may be approximated
by the magnetic dipole braking\footnote{This component may actually be some form of
particle outflow.}(Shapiro \& Teukolsky 1983)
\begin{equation}
 \dot{E}_{\rm d} = \frac{2\mu^2 \Omega^4}{3 c^3} \sin^2\alpha,
\end{equation}
where $\mu$ is the magnetic dipole moment, $\Omega$ is the angular rotation rate of the neutron star,
$c$ is the speed of light, and $\alpha$ is the angle between the rotation axis and the magnetic axis
(i.e., magnetic inclination angle). The effect of parallel magnetic moment is associated with particle acceleration
(e.g., inner vacuum accelerator, Ruderman \& Sutherland 1975).
The rotational energy loss rate due to this component may be written as (i.e., particle wind, Xu \& Qiao 2001)
\begin{equation}\label{particle}
 \dot{E}_{\rm p} = 2\pi r_{\rm p}^2 c \rho \Delta \phi,
\end{equation}
where $r_{\rm p}$ is polar cap radius, $\rho$ is charge density in the acceleration region,
$\Delta \phi$ is corresponding acceleration potential. The polar cap radius is
$r_{\rm p} = R(R\Omega/c)^{1/2}$ (where $R$ is neutron star radius) if the large scale magnetic field geometry
is of dipole form. Observationally, the different on states of intermittent pulsars are quite stable
(i.e., no significant variations between different on states, Lorimer et al. 2012; Young et al. 2013).
One way to achieve this is that the charge density in the acceleration region is always
the same\footnote{Other more complicated/more interesting possibilities is also possible.}.
The most naturally
value for the charge density is the Goldreich-Julian density (Goldreich \& Julian 1969). Therefore,
we chose $\rho =\rho_{\rm GJ} = \Omega B/(2\pi c)$, where $B$ is polar magnetic field. The polar magnetic
field is related with the magnetic moment as $\mu =1/2 B R^3$ (Shapiro \& Teukolsky 1983).
The presence of acceleration potential can accelerate primary particles. Secondary particles are generated
subsequently\footnote{The density of secondary
particles can be much higher than the Goldreich-Julian density, since they can have a much larger
multiplicity.}. These particles are responsible for the radio (and high energy) emissions of pulsars.
Meanwhile, they will also contribute to the rotational energy loss rate of the central neutron star.
There are various proposals for the acceleration potential (see Xu \& Qiao 2001 and references therein).
Different acceleration potential results in different rotational energy loss rate.
Therefore, pulsar timing observations can be employed to distinguish between different particle acceleration models.

The maximum acceleration potential for a rotating dipole is (Ruderman \& Sutherland 1975)
$\Delta \Phi = \mu \Omega^2/c^2$. Therefore, the rotational energy loss rate due to the particle wind
can be rewritten as
\begin{equation}
 \dot{E}_{\rm p} =\frac{2\mu^2 \Omega^4}{3 c^3} 3 \frac{\Delta \phi}{\Delta \Phi}.
\end{equation}
The total rotational energy loss rate is the combination of the perpendicular component and the parallel component.
Assuming the magnetic dipole component and the particle wind component contribute independently, the total
rotational energy loss rate is
\begin{equation}\label{Case II}
 \dot{E} = \dot{E}_{\rm d} + \dot{E}_{\rm p} =\frac{2\mu^2 \Omega^4}{3 c^3} (\sin^2\alpha
 + 3 \frac{\Delta \phi}{\Delta \Phi}), \quad \mbox{Case II}.
\end{equation}
Noting that the particle component may be mainly related with the parallel magnetic moment. The parallel
magnetic moment is $\mu_{||} = \mu \cos\alpha$. Taking this point into consideration, the corresponding
rotational energy loss rate is (Xu \& Qiao 2001)
\begin{equation}\label{Case I}
 \dot{E} = \frac{2\mu^2 \Omega^4}{3 c^3} (\sin^2\alpha
 + 3 \frac{\Delta \phi}{\Delta \Phi} \cos^2\alpha), \quad \mbox{Case I}.
\end{equation}
The above two cases are named case I and case II, respectively (according to the time order when
they first appear in literature). Case I was originally
considered by Xu \& Qiao (2001). Similar expression was also obtained by Contoupolous \& Spitkovsky (2006).
Case II is also possible considering recent numerical simulations
(Spitkovsky 2006; Li et al. 2012a). These two cases only differ by a factor of $\cos^2\alpha$.

The total rotational energy loss rate in the presence of a particle wind can  be rewritten as (Xu \& Qiao 2001)
\begin{equation}
 \dot{E} =\frac{2\mu^2 \Omega^4}{3 c^3} \eta,
\end{equation}
with $\eta=\sin^2\alpha + 3\Delta\phi/\Delta\Phi \, \cos^2\alpha$ (case I). Given the acceleration potential $\Delta \phi$,
$\eta$ and the rotational energy loss rate are known. The simplest case may be a constant acceleration potential (Yue et al. 2007).
If acceleration potential $\Delta \phi=3\times 10^{12} \,\rm Volts$,
then $\eta =\sin^2\alpha+ 54 B_{12}^{-1} \Omega^{-2} \cos^2\alpha$
(where $B_{12}$ is polar magnetic field in units of $10^{12} \,\rm G$).
Various physically motivated particle acceleration models are proposed since Ruderman \& Sutherland (1975).
The expression of $\eta$ for various acceleration models are listed in Table \ref{tab_eta}. The corresponding expression for case II can be
obtained by dropping the factor $\cos^2\alpha$.
Xu \& Qiao (2001) described six pulsar emission models, i.e., two vacuum gap (VG) models,
three space charge limited flow (SCLF) models and the outer gap (OG) model. In the vacuum gap model (Ruderman \&
Sutherland 1975), the situation that primary electrons emit $\gamma$-rays via curvature
radiation is named CR-induced and the one that primary electrons emit $\gamma$-rays via
resonant inverse Compton scattering off the thermal photons (e.g., Zhang et al. 2000)
is named ICS-induced. For the SCLF model (e.g., Arons \& Scharlemann 1979; Harding \&
Muslimov 1998), regimes I and II are considered, which are respectively defined as
the extreme cases without or with field saturation. Three models are included, i.e., the
CR-induced SCLF model in regime II, the ICS-induced SCLF model in regime II and the SCLF
model in regime I. The potential drop, which is model dependent, is different in each model,
so is the $\eta$-value.
The following calculations will be mainly for
case I. The calculation in case II is similar and will be discussed later.

 \begin{table}[!htbp]
 \centering
 \caption{Expressions of $\eta$ for seven particle acceleration models (mainly from Xu \& Qiao 2001 and references therein).
 The neutron star radius is taken as $10^6 \,\rm cm$.}
 \label{tab_eta}
    \begin{tabular}{lll}
      \hline\hline
     No. & Acceleration model & $\eta$ \\\hline
     1 &  VG (CR) &  $\sin^2\alpha+ 4.96\times 10^2 B_{12}^{-8/7} \Omega^{-15/7} \cos^2\alpha$\\
     2 & VG (ICS) & $\sin^2\alpha+ 1.02\times 10^5 B_{12}^{-22/7} \Omega^{-13/7} \cos^2\alpha$\\
     3 & SCLF (II,CR) & $\sin^2\alpha+ 38 B_{12}^{-1} \Omega^{-7/4} \cos^2\alpha$\\
     4 & SCLF (II, ICS) & $\sin^2\alpha+ 2.3 B_{12}^{-22/13} \Omega^{-8/13} \cos^2\alpha$\\
     5 & SCLF (I)$^a$ & $\sin^2\alpha+ 9.8\times 10^2 B_{12}^{-8/7} \Omega^{-15/7} \cos^2\alpha$\\
     6 &  OG$^b$ & $\sin^2\alpha+ 2.25\times 10^5 B_{12}^{-12/7} \Omega^{-26/7} \cos^2\alpha$\\
     7 & CAP$^c$ & $\sin^2\alpha+ 54 B_{12}^{-1} \Omega^{-2} \cos^2\alpha$\\
      \hline
    \end{tabular}
  \flushleft

$^a$: a factor $10^2$ is missed in Xu \& Qiao (2001).\\

$^b$: considering the modification due to Wu, Xu, \& Gil (2003). \\

$^c$: The case of constant acceleration potential (CAP, Yue et al. 2007).
The gap potential is assumed to be $\Delta \phi=3\times 10^{12} \,\rm Volts$.

   \end{table}

\section{Pulsar wind model for the spindown behavior of intermittent pulsars}

\subsection{Modeling intermittent pulsar on and off state}

If the on and off states of intermittent pulsars are solely due to a different geometry,
then it may be hard to explain the enhanced spindown during the on state.
A particle wind can contribute to both the radio emission and rotational energy
loss of the neutron star.
Then, it seems possible that, compared with the off state,
there is an additional particle wind during the on state of the intermittent pulsar.
A natural value for the density of this additional particle component
is the Goldreich-Julian density (Goldreich \& Julian 1969).
During the off state, there should also be some amount of particle outflow.
However, these particles failed somewhere during the acceleration
and subsequent radiation process. Therefore, there is no radio emission detected during
the off state (Lorimer et al. 2012). The rotational energy loss rate due to the
off state particle component must be smaller than that due to the on state particle component.
If the contribution from the off state particle component is small compared with the dipole radiation term,
then magnetic dipole radiation will dominate the rotational energy loss rate
in the off state of intermittent pulsars (Kramer et al. 2006).

If the off state magnetosphere of the intermittent pulsar is similar to the magnetosphere
of the pulsar in the death valley, the magnetic dipole approximation may be valid. Previous
studies showed that as the pulsar ages, the particle wind contribution to the
rotational energy loss rate will gradually disappear (Contoupolous \& Spitkovsky 2006; Tong \& Xu 2012).
When the pulsar passes through the death line, the remaining rotational energy loss rate is
identical to that of magnetic dipole radiation (e.g., proportional to $\sin^2\alpha$).
Numerical simulations are consistent with the analytical treatment (Li et al. 2012a).
However, it is also possible that the off state magnetosphere of the intermittent pulsar
is only a temporal failure (e.g., failure for one month or one year). In this case it may be different
from the magnetosphere of dead pulsars.
X-ray observations during the off state may tell us some details of this particle component
(Lorimer et al. 2012, with negative results, future observations will be more promising).

\subsection{Calculations in Case I (Equation (\ref{Case I}))}

The origin of both dipole radiation and particle wind is from the star's rotational energy: $\dot{E}=-I\Omega \dot{\Omega}$
($I$ is the moment of inertia, Lorimer \& Kramer 2005).
From the previous discussions, the spindown rate  of intermittent pulsars during the on and off state are respectively
\begin{eqnarray}
\label{Edot}
\dot{E} &=& -I \Omega \dot{\Omega}_{\rm on}\\
\label{Edotd}
\dot{E}_{\rm d} &=& -I \Omega \dot{\Omega}_{\rm off}.
\end{eqnarray}
The above equation can be rewritten as
\begin{eqnarray}
\label{on state}
\dot{\Omega}_{\rm on} &=& -\frac{2\mu^2}{3 I c^3}\Omega^3 \eta,\\
\label{off state}
\dot{\Omega}_{\rm off} &=& -\frac{2\mu^2}{3 I c^3}\Omega^3\sin^2\alpha.
\end{eqnarray}
Equation (\ref{off state}) can be rewritten as (assuming a moment of inertial of $10^{45}\,\rm g\, cm^2$)
\begin{equation}\label{equation1}
 B \sin\alpha= 6.4\times 10^{19} \sqrt{P \dot{P}_{\rm off}} \equiv B_{\rm c},
\end{equation}
where $P$ is the pulsar period, $\dot{P}_{\rm off}$ is period derivative during the off state,
$B_{\rm c}$ is defined as the characteristic magnetic field.
If the characteristic magnetic field is taken as the star's
true magnetic field, this corresponds to a inclination angle $\alpha=90^{\circ}$ (Kramer et al. 2006).
However, in the general case,
the magnetic field and the inclination angle should be solved simultaneously. This can be achieved
for intermittent pulsars since their spindown ratio during the on and off state can be measured.
Dividing Equation (\ref{on state}) by Equation (\ref{off state}), the spindown ratio of intermittent pulsars
between the on and the off state is
\begin{equation}\label{equation2}
r\equiv \frac{\dot{\Omega}_{\rm on}}{\dot{\Omega}_{\rm off}} = \frac {\eta}{\sin^2\alpha}
= \frac {\sin^2\alpha+3 (\Delta \phi/\Delta \Phi) \cos^2\alpha}{\sin^2\alpha}.,
\end{equation}

Solving Equation (\ref{equation1}) and (\ref{equation2}) simultaneously, using the
$\eta$ values listed in Table \ref{tab_eta}, the resulting magnetic inclination angles for each intermittent pulsar
in different acceleration models are listed in Table \ref{tab_inclination}.
The corresponding polar magnetic field can be directly obtained once the magnetic inclination
angle is known: $B=B_{\rm c}/\sin\alpha$ (from Equation (\ref{equation1})).
No solution exists for SCLF (II, ICS), since the corresponding magnetic inclination
is very small and the polar magnetic field is too high to be physically acceptable
(it is in the magnetar range). From the expression of $\eta$, for a specific inclination angle $\alpha$,
$\pi-\alpha$ gives the same result.
Table \ref{tab_inclination} only shows the case of $\alpha < 90^{\circ}$.

The calculation of braking index ($n\equiv \frac{\Omega \ddot{\Omega}}{\dot{\Omega}^2}$)
for intermittent pulsars is also straightforward.
From Equation (\ref{off state}), the braking index during the off state is exactly three,
if the moment of inertia, magnetic moment and inclination angle are constant.
While from Equation (\ref{on state}), the braking index during the on state
depends on the form of the acceleration potential. By differentiating Equation (\ref{on state}),
the corresponding braking index is (Xu \& Qiao 2001)
\begin{equation}\label{n}
n=3+\frac{\Omega}{\eta} \frac{{\rm d} \eta}{{\rm d} \Omega}.
\end{equation}
The general expression for $\eta$ is: $\eta = \sin^2\alpha + k \Omega^{-a} B_{12}^{-b}$
(see Table \ref{tab_eta}, where $k$ is constant). Therefore, the braking index for intermittent pulsars during the on state
is
\begin{equation}
 n_{\rm on} =3-\frac{r-1}{r} a.
\end{equation}
It only depends on the spindown ratio and the coefficient $a$
($a$ is determined by the acceleration potential dependence on $\Omega$). Acceleration potentials
with the same $\Omega$ dependence will have the same braking index. The braking index does not dependent on
the different combinations of magnetic dipole radiation and particle wind. Therefore, the braking index
is the same for case I and case II. The on state braking indices for the current three intermittent pulsars
are shown in Table \ref{tab_index}. The minimum braking index for each acceleration model is also shown\footnote{
The minimum braking index corresponds to when the particle wind dominates the rotational energy loss rate.
From the definition of braking index, the minimum braking index for a specific
acceleration model is $n_{\rm min} = 3-a$.}.
The SCLF (II, ICS) model has a minimum braking index of $2.4$. While the observed minimum braking index of pulsars
is $0.9\pm0.2$ (Esponiza et al. 2011). Therefore, from the braking index point of view, the SCLF (II, ICS)
model can also be ruled out.

\begin{figure}[!ht]
\centering
  \includegraphics[width=0.6\textwidth]{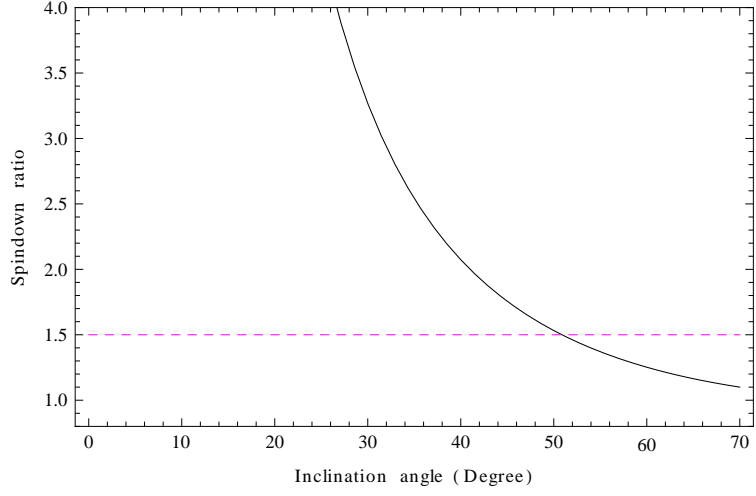}
  \caption{The spindown ratio of PSR B1931+24 as a function of inclination angle
           in the vacuum gap model induced by curvature radiation. The dashed line
           is the observed spindown ratio of PSR B1931+24 (Kramer et al. 2006).}
  \label{fig_ratio}
\end{figure}

\begin{figure}[!ht]
\centering
  \includegraphics[width=0.6\textwidth]{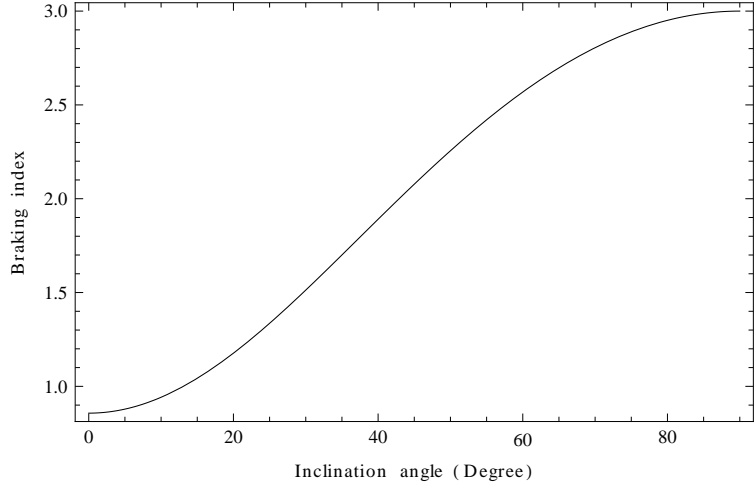}
  \caption{The braking index of PSR B1931+24 as a function of inclination angle
           in the vacuum gap model induced by curvature radiation.  }
  \label{fig_index}
\end{figure}

Figure \ref{fig_ratio} and \ref{fig_index} show separately
the spindown ratio and the braking index of PSR B1931$+$24 in the vacuum gap model
induced by curvature radiation. As the inclination angle decreases, the
spindown ratio between the on state and the off state  will be larger.
And the braking index gradually approaches one (the exact value depends on the acceleration potential).
This is because as the inclination angle decreases, the particle wind will gradually dominate
the star's rotational energy loss rate.
Other models for PSR B1931$+$24 and models for other intermittent pulsars (PSR J1841$-$0500 and PSR J1832$+$0029)
are different only in quantity.
Taking PSR B1931$+$24 as an example, the constant acceleration potential model gives the
smallest inclination angle $19^{\circ}$. The two biggest ones among them are
$80^{\circ}$  and $75^{\circ}$ , provided respectively by the VG (ICS)
model and the OG model. The theoretical range of braking index for this pulsar is from
1.7 to 2.4.

\begin{table}[!htbp]
  \centering
  \caption{Inclination angle of the three intermittent pulsars in different particle acceleration models,
  in units of degrees. No solution exists for SCLF (II, ICS).}
  \label{tab_inclination}

  \begin{tabular}{lccccccc}
    \hline\hline
     Pulsar name & $\rm VG(CR)$ & $\rm VG(ICS)$ & $\rm SCLF(II,CR)$ & $\rm SCLF(II,ICS)$ & $\rm SCLF(I)$ & $\rm OG$ & CAP\\\hline
    B1931+24 & 51 & 80 & 22 & - & 62 & 75 & 19 \\
    J1841$-$0500 & 24 & 53 & 6.7 & - & 38 & 60  & 6\\
    J1832+0029 & 53 & 87 & 24 & - & 63 & 76 & 19 \\
    \hline
  \end{tabular}
\end{table}

\begin{table}[!htbp]
  \centering
  \caption{On state braking index of the three intermittent pulsars in different particle acceleration models. The last row
  shows the minimum braking index for each model.}
  \label{tab_index}

  \begin{tabular}{lccccccc}
    \hline\hline
     Pulsar Name & $\rm VG(CR)$ & $\rm VG(ICS)$ & $\rm SCLF(II,CR)$ & $\rm SCLF(II,ICS)^{\ast}$ & $\rm SCLF(I)$ & $\rm OG$ & CAP
        \\\hline
    B1931+24 & 2.3 & 2.4 & 2.4 & - & 2.3 & 1.7 & 2.3 \\
    J1841$-$0500 & 1.7 & 1.9 & 2.0 & - & 1.7 & 0.77 & 1.8\\
    J1832+0029 & 2.1 & 2.2 & 2.2 & - & 2.1 & 1.4 & 2.1\\
         \hline
    $n_{\rm min}$ & 0.86 & 1.1 & 1.3 & 2.4 & 0.86 & $-0.71$ & 1\\
    \hline
  \end{tabular}
\flushleft
$^\ast$: SCLF (II, ICS) can already be ruled out.
\end{table}

\subsection{Discussions of Case II (Equation (\ref{Case II}))}

Case II differs from case I quantitatively. When obtaining the spindown ratio of intermittent
pulsars (Equation (\ref{equation2})), many unknown factors are concealed (e.g. the moment of
inertia, the magnetic dipole moment). And the observed spindown ratio is just a factor of two
(from 1.5 to 2.5). Therefore, in modeling the spindown ratio of intermittent pulsars,
the rotational energy loss rate has to be accurate within a factor of two (relative to magnetic dipole braking case).
For case II, Equation (\ref{equation1}) is still valid. The expression for the spindown ratio in case II
is
\begin{equation}\label{equation2_caseII}
 r\equiv \frac{\dot{\Omega}_{\rm on}}{\dot{\Omega}_{\rm off}} = \frac{\eta}{\sin^2\alpha}
 = \frac {\sin^2\alpha+3 \Delta \phi/\Delta \Phi}{\sin^2\alpha}.
\end{equation}
Generally, $\eta= \sin^2\alpha + k \Omega^{-a} B_{12}^{-b}$ (see Table \ref{tab_eta}.
For case II, $k$ does not include $\cos^2\alpha$.).
Using Equation (\ref{equation1}), Equation (\ref{equation2_caseII}) can be rewritten as
\begin{equation}
 r=1+ \frac{k\Omega^{-a} B_{12}^{2-b}}{B_{\rm c,12}^2},
\end{equation}
where $B_{\rm c,12}$ is the characteristic magnetic field in units of $10^{12} \,\rm G$.
When $2-b>0$, and considering the polar magnetic field is always larger than or equal
to the characteristic magnetic field, there is a lower limit on the spindown
ratio in each acceleration model
\begin{equation}
 r \ge 1+ k \Omega^{-a} B_{\rm c, 12}^{-b}.
\end{equation}
Observationally, PSR B1931+24 has the smallest spindown ratio: $r=1.5$ (Kramer et al. 2006). Acceleration potentials with
their minimum spindown ratio larger than 1.5 can be ruled out. At the same time, if the corresponding magnetic
inclination angle is very small and the polar magnetic field is too large, then the acceleration potential
is also not favored. Only two models meet these two
criteria: SCLF (II, CR) and the constant acceleration potential model.
The magnetic inclination angle in case II is similar to the corresponding
values in case I (at most a few degrees larger).

\subsection{Duty cycle of particle wind determined from timing observations}

The particle wind component may only work for part of the time (Harding et al. 1999;
Tong et al. 2013). Denote the duty cycle of particle wind as $D_{\rm p}$ (fractional time when the particle wind is present),
the long term averaged spindown of the pulsar is (Harding et al. 1999)
\begin{equation}\label{Dp}
 -I \Omega \dot{\Omega}_{\rm ave} = \dot{E}_{\rm d} (1-D_{\rm p}) + \dot{E} D_{\rm p},
\end{equation}
where $\dot{\Omega}_{\rm ave}$ is the long term averaged spindown rate. For intermittent pulsars,
from Equation (\ref{Edot}) and (\ref{Edotd}), Equation (\ref{Dp}) can be rewritten as
\begin{equation}
 \dot{\Omega}_{\rm ave} = \dot{\Omega}_{\rm off} (1-D_{\rm p}) + \dot{\Omega}_{\rm on} D_{\rm p}.
\end{equation}
Therefore, the duty cycle of particle wind can be determined from timing observations
\begin{equation}\label{Dp2}
 D_{\rm p} = \frac{\dot{\Omega}_{\rm ave} - \dot{\Omega}_{\rm off}}{\dot{\Omega}_{\rm on} - \dot{\Omega}_{\rm off}}
 = \frac{\dot{\nu}_{\rm ave} - \dot{\nu}_{\rm off}}{\dot{\nu}_{\rm on} - \dot{\nu}_{\rm off}}.
\end{equation}
At present, only PSR B1931+24 had $\dot{\nu}_{\rm on}$, $\dot{\nu}_{\rm off}$ and $\dot{\nu}_{\rm ave}$ reported (Kramer et al. 2006).
From the long term averaged spindown, the duty cycle of its particle wind is $D_{\rm p} =26\%$.
In the pulsar wind model, the presence of particle wind corresponds to the on state. Then the duty
cycle of particle wind should be the same as the duty cycle of the on state.
Kramer et al. (2006) estimated that
the duty cycle of the on state is  $\sim 20\%$. Later study gave a duty cycle of the on state $26\pm 6\%$
(Young et al. 2013). The two duty cycles, one obtained from long term averaged spindown (Equation (\ref{Dp2}))
and one from radiation monitoring, are consistent with each other.
We suggest future timing observations of intermittent pulsars give not only its spindown rate
during the on/off state ($\dot{\nu}_{\rm on}$ and $\dot{\nu}_{\rm off}$) but also the long term averaged
spindown rate ($\dot{\nu}_{\rm ave}$).

\section{Discussions and conclusions}

In pulsar researches, the magnetic dipole braking assumption is often employed, e.g., in calculating the
characteristic magnetic field and the characteristic age.
The magnetic dipole braking assumes a rotating perpendicular dipole in vacuum. A real pulsar must
have a magnetosphere (Goldreich \& Julian 1969). Pulsars can radiate radio and high energy photons.
Therefore, in the pulsar magnetosphere there must be some kind of particle acceleration and radiation. Previous studies showed
that the magnetic dipole braking assumption is correct to the lowest order approximation (Goldreich \& Julian 1969; Xu \& Qiao 2001;
Contoupolous \& Spitkovsky 2006). When considering higher order effects, e.g. braking index, timing noise,
and intermittent pulsars, physically based models must be employed. Therefore,
the magnetic dipole braking model is just a pedagogical model (Shapiro \& Teukolsky 1983). The characteristic magnetic
field is just the effective magnetic field (all the torque terms are attributed to a perpendicular dipole field).


Applying the pulsar wind model of Xu \& Qiao (2001)
to the three intermittent pulsars, we calculated their corresponding inclination angles and
braking indices. Table \ref{tab_inclination} and \ref{tab_index} show that the results are quite
different from each other. It is because they have different acceleration potentials.
In modeling the particle wind component, only one kind of acceleration potential is considered.
The real case may allow the coexistence of different acceleration potentials (e.g., a core gap and
a more extended one, Qiao et al. 2007; Du et al. 2010). When intermittent pulsars have more observations,
it will be necessary to check the coexistence of different acceleration potentials.
Considering the seven acceleration potentials in Table \ref{tab_inclination} and Table \ref{tab_index}
(one of them can already be ruled out),
we suggest that the curvature induced vacuum gap model (VG (CR) model, Ruderman \& Sutherland 1975) be
compared with observations first\footnote{Since it was among the first pulsar models proposed and is still user-friendly.
However, for normal pulsars with non-magnetar strength field,
space charge limited flow models are favored, unless the central star is a bare quark star (Yu \& Xu 2011 and references therein).}.

From the above calculations, the inverse Compton scattering induced space charge limited flow with field saturation
(i.e., the SCLF (II, ICS) model) can already be ruled out.
The corresponding magnetic inclination angle is too small and the polar magnetic
field is in the magnetar range. On the other hand, this may actually happen in some magnetars.
Magnetars may be wind braking and some of them may have a small inclination angle and a very high
dipole magnetic field (Tong \& Xu 2012; Tong et al. 2013). For PSR J1841$-$0500, its inclination
angle is also relatively small in the constant acceleration potential (CAP) case. This
depends on the gap potential adopted ($3\times 10^{12} \,\rm Volts$ at present). If the
gap potential is $10^{13} \,\rm Volts$, then the consequent inclination angle will be $18$ degrees.

The spindown behavior of intermittent pulsars looks like a glitch in the interval between
successive on states (e.g., Figure 1 in Camilo et al. 2012). However, a more physically motivated explanation
is that the absence of particle outflow is responsible for both the cessation of radio emission
and the lower slow down rate during the off state (Kramer et al. 2006; and this paper). The opposite case
is also possible: during the observational interval, the particle wind may be significantly stronger.
Then the spindown behavior of the pulsar will look like a negative glitch.
Such negative glitch has already been observed in one magnetar (i.e., anti-glitch, Archibald et al. 2013).
A stronger particle wind during the observational interval can explain the anti-glitch (Tong 2014).
Therefore, the spindown behavior of intermittent pulsars and anti-glitch can be understood
uniformly in the wind braking scenario.

The effective model proposed by Kramer et al. (2006) explains the different spindown ratio
of intermittent pulsars by a different particle density.
PSR J1841$-$0500 has the largest spindown ratio ($r=2.5$). Adopting the prescription of Kramer et al. (2006),
the required plasma density is three times the Goldreich-Julian density (Lorimer et al. 2012).
In our model, the charge density in the acceleration region (i.e., primary particles) can always be the
Goldreich-Julian density. A constant amount of charged particles accelerated
may ensure that different on states of intermittent pulsars are stable (Young et al. 2013).
Kramer et al. (2006) employed the pulsar wind model of Harding et al. (1999)
to treat the wind torque.
It is equivalent to the case that the
accelerated particles can attain the maximum potential (Tong et al. 2013).
In the real case, the acceleration gap must break down somewhere due to pair screening
(Ruderman \& Sutherland 1975).
It also assumed an orthogonal rotator when
calculating the surface dipole field using the off state spindown rate.
In our pulsar wind model, physically motivated particle acceleration potentials are taken
into consideration. For a specific particle acceleration potential, the spindown ratio is determined by
the magnetic inclination angle (see Figure \ref{fig_ratio}). 	
The surface dipole field and the magnetic inclination angle are solved simultaneously.

In the pulsar wind model considered in this paper,
there is a particle wind torque in addition to the magnetic dipole
torque during the on state. Therefore, the spindown rate should always be larger in the on state
than in the off state. Previous literatures also discussed the possibility of no dipole radiation during
the on state (e.g., Geurevich \& Istomin 2007). From Equation (\ref{equation2}),
if there is no dipole term in the on state, the spindown ratio will be proportional to
$\cot^2\alpha$ (see also Geurevich \& Istomin 2007; Beskin \& Nokhrina 2007).
For large inclination angle, the spindown ratio can be smaller than one.
Some intermittent pulsars are expected to have smaller spindown rate in the on state than in the off state.
However, the present three sources all have larger spindown rates during the on state.
Therefore, it is more likely that the dipole term coexists with the particle wind term
in the on state of intermittent pulsars (and normal pulsars).

Li et al. (2012b) modeled the spindown of intermittent pulsars using different states of
pulsar magnetosphere. Their work is from the magnetohydrodynamical simulation point of view.
The current of charged particles is modeled by introducing some dissipation in the magnetosphere.
The Li et al. (2012b) model can be viewed as macroscopic version of particle wind.
While this work treats the particle wind from the microscopic point of view.
The spindown ratio also depends on the magnetic inclination angle in Li et al. (2012b).
Li et al. (2012b) and this work predict quantitatively different inclination angles
(see Figure 4 there and Figure 1 in this paper).
Our pulsar wind model can predict the braking index of the intermittent pulsar
 (with braking index between one and three).

Since intermittent pulsars are very weak, therefore large telescope (e.g.,
Arecibo and the Five hundred meter Aperture Spherical radio Telescope, known as FAST)
is required to make polarization observations.
In the future, FAST pulsar survey may discover more intermittent pulsars (Nan et al. 2011).
Polarization observations of more
intermittent pulsars will help us to test different models of pulsar magnetospheres.
In combination with Green Bank Telescope (and other large telescopes),
the braking index of intermittent pulsars (the present three sources
and more ones to be discovered) may be measured. For one intermittent pulsar, if the spindown
ratio, the magnetic inclination angle, and the braking index are all measured,
they will provide very strong constraint on our understanding of pulsar
magnetospheres.

In summary, the pulsar wind model of Xu \& Qiao (2001) is employed to explain the spindown behavior of intermittent pulsars.
The enhanced spindown during the on state is due to the presence of an additional particle wind.
In modeling the pulsar wind, the effect of particle acceleration is included.
Using the measured spindown ratio, the magnetic field and the inclination angle of intermittent pulsars
are solved simultaneously. Their predicted braking indices are also shown.
The density of the accelerated particles can always equal to the Goldreich-Julian density.
This may guarantee that different on states of intermittent pulsars are very stable.
The duty cycle of particle wind can be determined from the long term averaged spindown.
It is consistent with the cycle of the on state.

\section*{Acknowledgments}
The authors would like to thank the referee very much for helpful comments,
and P.F.Wang for discussions. L.Li is supported by Natural Science Foundation of Xinjing University (07.02.0432.02),
H.Tong is supported by NSFC (11103021), WLFC (LHXZ201201), Xinjiang Bairen project,
and Qing Cu Hui of CAS. W. M. Yan is supported by NSFC (11203063) and WLFC (XBBS201123).
J. P. Yuan is supported by NSFC (11173041) and WLFC (XBBS201021)).
R.X.Xu is supported by National Basic Research Program of China (2012CB821800),
and NSFC (11225314).


\label{lastpage}

\end{document}